\documentclass[prl,aps,twocolumn]{revtex4}

\usepackage{psfrag}
\usepackage{graphicx}
\usepackage{dcolumn}
\usepackage{latexsym,amsfonts}
\usepackage{bm}
\usepackage{amssymb}
\begin{document}

\title{On Continuum- and Bound-State $\ell^-$--Decay Rates of Pionic
  and Kaonic Hydrogen in the Ground State}

\author{M. Faber$^{a}$\thanks{E-mail: faber@kph.tuwien.ac.at},
A. N. Ivanov${^{a,b}}$, V. A. Ivanova$^{c}$, J. Marton${^{b}}$
M. Pitschmann${^a}$, N. I. Troitskaya$^{c}$, M. Wellenzohn$^{a}$}
\affiliation{${^a}$Atominstitut der \"Osterreichischen
Universit\"aten, Technische Universit\"at Wien, Wiedner Hauptstrasse
8-10, A-1040 Wien, Austria} \affiliation{${^b}$Stefan Meyer Institut
f\"ur subatomare Physik \"Osterreichische Akademie der Wissenschaften,
Boltzmanngasse 3, A-1090, Wien, Austria}\affiliation{ $^e$ State
Polytechnic University of St. Petersburg, Polytechnicheskaya 29,
195251, Russian Federation} \email{ivanov@kph.tuwien.ac.at}

\date{\today}

\begin{abstract}
We calculate the continuum- and bound-state $\ell^-$--decay rates of
pionic and kaonic hydrogen in the ground state, where $\ell^-$ is
either the electron $e^-$ or the muon $\mu^-$.  
\end{abstract}

\maketitle

\section{Introduction}

It is well--known that pionic and kaonic hydrogen in the ground state
play an important role for the investigation of low--energy strong
interactions \cite{SMI}. Pionic and kaonic hydrogen are the bound
states, where the electron is replaced by a negatively charged
$\pi^-$--meson and $K^-$--meson, respectively. Since the $\pi^-$ and
$K^-$ mesons are unstable under weak decays into the leptonic pairs
$\ell^- \tilde{\nu}_{\ell}$, where $\ell^-$ is either the electron
$e^-$ or muon $\mu^-$ and $\tilde{\nu}_{\ell}$ is the antineutrino
with the electron $\tilde{\nu}_e$ or muon $\tilde{\nu}_{\mu}$ flavour,
the lifetimes of mesic hydrogens should be restricted by the lifetimes
of mesons.  The lifetimes of the free $\pi^-$ and $K^-$ mesons are
equal to $\tau_{\pi^-} = 2.60\times 10^{-8}\,{\rm s}$ and $\tau_{K^-}
= 1.24\times 10^{-8}\,{\rm s}$, respectively \cite{PDG06}. However,
one can assume that these lifetimes do not define real lifetimes of
pionic and kaonic hydrogen in the ground state.  Since a negatively
charged meson is bound, this can probably change the lifetime of mesic
hydrogen. In addition mesic hydrogen can have also the decay channels
different to the emission of the free leptonic pairs $\ell^-
\tilde{\nu}_{\ell}$ only. Indeed, mesic hydrogen, which we denote as
${\rm H}^{(1s)}_m$ for $m = \pi^-$ or $K^-$, is unstable under the
continuum--state $\ell^-$--decay ${\rm H}^{(1s)}_m \to p + \ell^- +
\tilde{\nu}_{\ell}$ with the emission of the free leptonic pair
$\ell^- \tilde{\nu}_{\ell}$ and the bound-state $\ell^-$--decay ${\rm
H}^{(1s)}_m \to {\rm H}_{\ell} + \tilde{\nu}_{\ell}$, where ${\rm
H}_{\ell}$ is hydrogen with a bound lepton $\ell^-$.

In this letter we calculate the continuum- and bound--state
$\ell^-$--decay rates of mesic hydrogen in the ground state. According
to the classification of the $\beta$--decays \cite{EK66,HS66}, these
are allowed $\ell^-$--decays with the selection rule $\Delta J^P =
0^+$.

The calculation of the continuum- and bound-state $\ell^-$--decay
rates of mesic hydrogen in the ground state we carry out in the
standard theory of weak interactions of hadrons \cite{STW} using the
technique developed in \cite{Ivanov1,Faber1} for the analysis of the
decay rates of the H--like and He--like heavy ions. The weak
interaction Hamilton density operator we take in the form \cite{STW}
\begin{eqnarray}\label{label1}
\hspace{-0.3in}{\cal H}_W(x) &=& \frac{G_F}{\sqrt{2}}\,V_{uq}\,J^{a +
ib}_{\mu}(x)\nonumber\\
\hspace{-0.3in}&&\times\,[\bar{\psi}_{\ell}(x)\gamma^{\mu}(1 -
\gamma^5)\psi_{\nu_{\ell}}(x)],
\end{eqnarray}
where $G_F = 1.166\times 10^{-11}\,{\rm MeV}^{-2}$ is Fermi's weak
constant, $V_{uq}$ is an element of the CKM--matrix with $q = d$ or
$s$ for the $\pi^-$ or $K^-$ meson, respectively, equal to $V_{ud} =
0.97377$ and $V_{us} = 0.22570$ \cite{PDG06}, $J^{a + ib}_{\mu}(x)$ is
a hadronic current with $V-A$ structure and $SU_f(3)$--flavour indices
$a = 1,4$ and $b = 2,5$ for the $\pi^-$ and $K^-$ meson decays,
respectively, $\psi_{\ell}(x)$ and $\psi_{\nu_{\ell}}(x)$ are the
field operators of the $\ell^-$--lepton and the anti--neutrino
(neutrino) with the $\ell$--lepton flavour.

The wave function of mesic hydrogen in the ground state we take in the
following form \cite{IV3}
\begin{eqnarray}\label{label2}
\hspace{-0.3in}&&|{\rm H}^{(1s)}_m(\vec{K},\sigma)\rangle =
\frac{1}{(2\pi)^3}\,\sqrt{2 E_{{\rm H}^{(1s)}_m}(\vec{K}\,)}\nonumber\\
\hspace{-0.3in}&&\times\int \frac{d^3k}{\sqrt{2
E_m(\vec{k}\,)}}\frac{d^3p}{\sqrt{2
E_p(\vec{p}\,)}}\,\delta^{(3)}(\vec{K} - \vec{k} -
\vec{p}\,)\nonumber\\
\hspace{-0.3in}&&\times\,\Phi_{1s}\Big(\frac{m_m \vec{p} - m_p
\vec{k}}{m_p +
m_m}\Big)\,c^{\dagger}_m(\vec{k}\,)a^{\dagger}_p(\vec{p},
\sigma)|0\rangle,
\end{eqnarray}
where $\Phi_{1s}(\vec{k}\,)$ is the wave function of mesic hydrogen in
the $1s$ ground state in the momentum representation \cite{BS57},
$c^{\dagger}_m(\vec{k}\,)$ and $a^{\dagger}_p(\vec{p}, \sigma)$ are
creation operators of the meson $m$ and the proton, respectively. The
operators of creation and annihilation of mesons and protons obey
standard relativistic covariant commutation and anti--commutation
relations, respectively \cite{IV3}. The energy of mesic hydrogen is
defined by $E_{{\rm H}^{(1s)}_m}(\vec{K}\,) = \sqrt{\vec{K}^{\,2} +
M^2_{{\rm H}^{(1s)}_m}}$, where $M_{{\rm H}^{(1s)}_m} = m_p + m_m +
\epsilon_{1s}$ and $\epsilon_{1s} = - \alpha/2 a_B = -
\alpha^2\mu_m/2$ are the mass of mesic hydrogen and the binding
energy, $\mu_m = m_m m_p/(m_m + m_p)$ is the reduced mass of the $mp$
pair and $a_B = 1/\alpha \mu_m$ is the Bohr radius.  The wave function
$\Phi_{1s}(\vec{k}\,)$ is equal to \cite{BS57}
\begin{eqnarray}\label{label3}
\Phi_{1s}(\vec{k}\,) = \frac{8\sqrt{\pi a^3_B}}{(1 + a^2_B k^2)^2}.
\end{eqnarray}
Since the wave function $\Phi_{1s}(\vec{k}\,)$ is normalised to unity
\cite{BS57}
\begin{eqnarray}\label{label4}
\int \frac{d^3k}{(2\pi)^3}\,|\Phi_{1s}(\vec{k}\,)|^2 = 1,
\end{eqnarray}
the wave function Eq.(\ref{label2}) has a standard relativistic
invariant normalisation \cite{IV3}
\begin{eqnarray}\label{label5}
\hspace{-0.3in}&&\langle {\rm H}^{(1s)}_m(\vec{K}\,',\sigma\,'\,)|{\rm
H}^{(1s)}_m(\vec{K},\sigma)\rangle = \nonumber\\
\hspace{-0.3in}&& = (2\pi)^3 2 E_{{\rm
H}^{(1s)}_m}(\vec{K}\,)\,\delta^{(3)}(\vec{K}\,' -
\vec{K}\,)\,\delta_{\sigma '\sigma}.
\end{eqnarray}
Using these definitions we can proceed to calculating the continuum-
and bound-state $\ell^-$--decay rates.

\section{Continuum-state $\ell^-$--decay rate of mesic hydrogen
in the ground state}

The $\mathbb{T}$--matrix of weak interactions, taken to first order in
perturbation theory, is equal to \cite{STW}
\begin{eqnarray}\label{label6}
\mathbb{T} = -\int d^4x\,{\cal H}_W(x)
\end{eqnarray}
For the ${\rm H}^{(1s)}_m \to p + \ell^- + \tilde{\nu}_{\ell}$ decay
the matrix element of the $\mathbb{T}$--matrix is
\begin{eqnarray}\label{label7}
\hspace{-0.3in}&&\langle \tilde{\nu}_{\ell}\ell^-p|\mathbb{T}|{\rm
H}^{(1s)}_m\rangle = (2\pi)^4 \delta^{(4)}(k_{\nu} + k_{\ell} + k_p -
K)\nonumber\\
\hspace{-0.3in}&&\hspace{0.97in}\times\,M({\rm H}^{(1s)}_m\to p \ell^-
\tilde{\nu}_{\ell}),
\end{eqnarray}
where $k_{\nu} = (E_{\nu}, \vec{k}_{\nu})$, $k_{\ell} = (E_{\ell},
\vec{k}_{\ell})$ , $k_p = (E_p, \vec{k}_p)$ and $K = (E_{{\rm
H}^{(1s)}_m},\vec{K}\,)$ are the 4-momenta of the anti--neutrino, the
lepton $\ell^-$, the proton and mesic hydrogen, respectively. The
lepton $\ell^-$, the proton and mesic hydrogen have the polarisations
$\sigma_{\ell}$, $\sigma_p$ and $\sigma$, respectively, the
anti--neutrino is polarised along its 3--momentum $\vec{k}_{\nu}$ with
$\sigma_{\nu} = + \frac{1}{2}$ \cite{STW}. The amplitude of
the ${\rm H}^{(1s)}_m \to p + \ell^- + \tilde{\nu}_{\ell}$ decay is
defined by
\begin{eqnarray}\label{label8}
M({\rm H}^{(1s)}_m \to p \ell^-\tilde{\nu}_{\ell}) = -\langle
\tilde{\nu}_{\ell}\ell^-p|{\cal H}_W(0)|{\rm H}^{(1s)}_m\rangle.
\end{eqnarray}
The matrix element $\langle \tilde{\nu}_{\ell}\ell^-p|{\cal
H}_W(0)|{\rm H}^{(1s)}_m\rangle$, which we calculate in the center of
mass frame of mesic hydrogen, takes the form of the product of the
leptonic current $[\bar{u}_{\ell}\gamma^{\mu}(1 -
\gamma^5)\,v_{\tilde{\nu}_{\ell}}]$, where $\bar{u}_{\ell}$ and
$v_{\tilde{\nu}_{\ell}}$ are Dirac bispinors of the lepton $\ell^-$
and the anti--neutrino $\tilde{\nu}_{\ell}$, respectively, and the
matrix element of the hadronic current $\langle
p|J^{a+ib}_{\mu}(0)|mp\rangle$.  The $\langle
p|J^{a+ib}_{\mu}(0)|mp\rangle$ contains {\it disconnected} and {\it
connected} parts
\begin{eqnarray}\label{label9}
\hspace{-0.3in}&&\langle
p(\vec{k}_p,\sigma_p)|J^{a+ib}_{\mu}(0)|m(\vec{k}\,)p(\vec{q},\sigma)\rangle
=\nonumber\\
\hspace{-0.3in}&&= \langle
p(\vec{k}_p,\sigma_p)|J^{a+ib}_{\mu}(0)|m(\vec{k}\,)p(\vec{q},\sigma)\rangle_{\rm
disconn.}\nonumber\\
\hspace{-0.3in}&&+\langle
p(\vec{k}_p,\sigma_p)|J^{a+ib}_{\mu}(0)|m(\vec{k}\,)p(\vec{q},\sigma)\rangle_{\rm
conn.}
\end{eqnarray}
The {\it disconnected} part takes the form
\begin{eqnarray}\label{label10}
\hspace{-0.3in}&&\langle
p(\vec{k}_p,\sigma_p)|J^{a+ib}_{\mu}(0)|m(\vec{k}\,)p(\vec{q},\sigma)\rangle_{\rm
disconn.} = - \delta_{\sigma_p \sigma}\nonumber\\
\hspace{-0.3in}&&\times\,(2\pi)^3 2E_p(\vec{q}\,)\delta^{(3)}(\vec{q}
- \vec{k}_p) \langle 0|A^{a + ib}_{\mu}(0)|m(\vec{k}\,)\rangle
\end{eqnarray}
and gives the main contribution to the continuum-state $\ell^-$-decay
rate.  The contribution of the {\it connected} part, which we define
in Appendix A, is smaller compared to the {\it disconnected} one.

As a result the amplitude Eq.(\ref{label8}), defined by the {\it
disconnected} part of the matrix element of the hadronic current, is
\cite{Ivanov1,IV3}
\begin{eqnarray}\label{label11}
\hspace{-0.3in}&&M({\rm H}^{(1s)}_m \to p \ell^-\tilde{\nu}_{\ell}) =
\delta_{\sigma_p\sigma}\,\sqrt{\frac{2 M_{{\rm H}^{(1s)}_m} 2
E_p(\vec{k}_p)}{2 E_m(\vec{k}_p)}}\nonumber\\
\hspace{-0.3in}&&\times\,\Phi_{1s}(\vec{k}_p)\,
\frac{G_F}{\sqrt{2}}\,V_{uq}\,\langle 0|A^{a +
ib}_{\mu}(0)|m(-\vec{k}_p)\rangle\nonumber\\
\hspace{-0.3in}&&\times\,
[\bar{u}_{\ell}(\vec{k}_{\ell},\sigma_{\ell})\gamma^{\mu}(1 -
\gamma^5)\,v_{\tilde{\nu}_{\ell}}(\vec{k}_{\nu},+ \frac{1}{2})].
\end{eqnarray}
The matrix element part $\langle 0|A^{a +
ib}_{\mu}(0)|m(-\vec{k}_p)\rangle$ we take in the standard form
\cite{PDG06} (see also \cite{STW,CA68})
\begin{eqnarray}\label{label12}
\langle 0|A^{a +
ib}_{\mu}(0)|m(-\vec{k}_p)\rangle = i\,\sqrt{2}\,F_m\,Q_{\mu},
\end{eqnarray}
where $Q = (E_m, -\,\vec{k}_p)$ with $E_m = \sqrt{\vec{k}^{\,2}_p +
M^2_m}$ and $M_m = m_m + \epsilon_{1s}$, $F_m$ is the PCAC constant of
the $m$--meson equal to $F_{\pi} = 92.4\,{\rm MeV}$ and $F_K =
113.0\,{\rm MeV}$ for the $\pi^-$ and $K^-$ meson, respectively
\cite{PDG06}.

The decay rate of the continuum-state $\ell^-$--decay is defined by \cite{Faber1}
\begin{eqnarray}\label{label13}
\lambda^{(m)}_{\ell^-_c} &=& \frac{1}{2 M_{{\rm H}^{(1s)}_m}}\int\!\!\!
\frac{d^3k_p}{(2\pi)^3 2 E_p}\frac{d^3k_{\ell}}{(2\pi)^3 2
E_{\ell}}\frac{d^3k_{\nu}}{(2\pi)^3 2 E_{\nu}}\nonumber\\
&&\times\,(2\pi)^4\delta^{(4)}(k_{\nu} + k_{\ell} + k_p -
K)\nonumber\\
&&\times\,\overline{|M({\rm H}^{(1s)}_m \to p
\ell^-\tilde{\nu}_{\ell})|^2},
\end{eqnarray}
where $\overline{|M({\rm H}^{(1s)}_m \to p
\ell^-\tilde{\nu}_{\ell})|^2}$ is given by
\begin{eqnarray}\label{label14}
\hspace{-0.in}&&\overline{|M({\rm H}^{(1s)}_m \to p
\ell^-\tilde{\nu}_{\ell})|^2} = \frac{1}{2}\sum_{\sigma}
\sum_{\sigma_p}\sum_{\sigma_{\ell}}\nonumber\\
\hspace{-0.3in}&&\times|M({\rm H}^{(1s)}_m \to p
\ell^-\tilde{\nu}_{\ell})|^2 = \frac{2 M_{{\rm H}^{(1s)}_m} 2
E_p(\vec{k}_p)}{2 E_m(\vec{k}_p)}\nonumber\\
\hspace{-0.3in}&&\times\,16\, G^2_F|V_{uq}|^2F^2_m\,
|\Phi_{1s}(\vec{k}_p)|^2\nonumber\\
\hspace{-0.3in}&&\times\,\Big((Q\cdot k_{\ell})(Q\cdot
k_{\nu}) - \frac{1}{2}\,Q^2(k_{\ell}\cdot k_{\nu})\Big).
\end{eqnarray}
Substituting Eq.(\ref{label14}) into Eq.(\ref{label13}) we arrive at
the decay rate
\begin{eqnarray}\label{label15}
\hspace{-0.3in}&&\lambda^{(m)}_{\ell^-_c} = \int\!\!\!
\frac{d^3k_p}{(2\pi)^3 2 E_m}\frac{d^3k_{\ell}}{(2\pi)^3 2
E_{\ell}}\frac{d^3k_{\nu}}{(2\pi)^3 2
E_{\nu}}\,|\Phi_{1s}(\vec{k}_p)|^2\nonumber\\
\hspace{-0.3in}&&\times\,(2\pi)^4\delta^{(4)}(k_{\nu} + k_{\ell} + k_p
- K)\,16\, G^2_F|V_{uq}|^2F^2_m \nonumber\\
\hspace{-0.3in}&&\times\,\Big((Q\cdot
k_{\ell})(Q\cdot k_{\nu}) - \frac{1}{2}\,Q^2(k_{\ell}\cdot
k_{\nu})\Big).
\end{eqnarray}
For the integration over the phase volume of the lepton
$\ell^-\tilde{\nu}_{\ell}$ pair we use the formula
\begin{eqnarray}\label{label16}
\hspace{-0.3in}&& T^{\alpha\beta}(P) = \frac{1}{2}\int
\Big(k^{\alpha}_{\ell}k^{\beta}_{\nu} +
k^{\beta}_{\ell}k^{\alpha}_{\nu} - g^{\alpha\beta}(k_{\ell}\cdot
k_{\nu})\Big)\nonumber\\
\hspace{-0.3in}&&\times\,(2\pi)^4\delta^{(4)}(P - k_{\nu} -
k_{\ell})\frac{d^3k_{\ell}}{(2\pi)^3 2
E_{\ell}}\frac{d^3k_{\nu}}{(2\pi)^3 2 E_{\nu}}\nonumber\\
\hspace{-0.3in}&& =\Big[ \Big(1 + 2
\frac{m^2_{\ell}}{P^2}\Big)\,\frac{P^{\alpha}P^{\beta}}{P^2} - \Big(1
+
\frac{1}{2}\,\frac{m^2_{\ell}}{P^2}\Big)\,g^{\alpha\beta}\Big]\nonumber\\
\hspace{-0.3in}&&\times\,\frac{P^2}{48\pi}\,\Big(1 -
\frac{m^2_{\ell}}{P^2}\Big)^2,
\end{eqnarray}
where $P = K - k_p \simeq Q$ and $Q^2 \simeq M^2_m \simeq m^2_m$, as
the main contribution to the integral over $\vec{k}_p$ comes from the
region $|\vec{k}_p| \sim 1/a_B = \alpha \mu_m$.

Substituting Eq.(\ref{label16}) into Eq.(\ref{label15}) and
integrating over $\vec{k}_p$ we get
\begin{eqnarray}\label{label17}
\hspace{-0.5in}&&\lambda^{(m)}_{\ell^-_c} = G^2_F|V_{uq}|^2 F^2_m
m^2_{\ell}\,\frac{(M^2_m - m^2_{\ell})^2}{4\pi M^3_m},
\end{eqnarray}
where we have used the normalisation condition Eq.(\ref{label4}) for
the wave function $\Phi_{1s}(\vec{k}_p)$. At the neglect of the
binding energy $M_m = m_m + \epsilon_{1s} \simeq m_m$ the obtained
continuum-state $\ell^-$--decay rate coincides with the
$\ell^-$--decay rate of a free $m$--meson \cite{STW}.

\section{Bound-state $\ell^-$--decay rate of mesic hydrogen
in the ground state}

In the bound-state $\ell^-$--decay of mesic hydrogen ${\rm
H}^{(1s)}_m$, ${\rm H}^{(1s)}_m \to {\rm H}_{\ell} +
\tilde{\nu}_{\ell}$, we get hydrogen ${\rm H}_{\ell}$ with the lepton
$\ell^-$, which is practically in the bound $(ns)$ state
\cite{Faber1}, where $n$ is the {\it principal} quantum number, and
the anti--neutrino $\tilde{\nu}_{\ell}$.  Due to the hyperfine
interaction \cite{HFS1} hydrogen ${\rm H}_{\ell}$ can be in the
hyperfine $(ns)_F$ states with atomic spin $F = 0$ and $F = 1$
\cite{Faber1,HFS1}.  The contribution of the excited $nL$--states with
$L > 0$, where $L = 1, \ldots, n - 1$ is the angular momentum, is
negligible small. Nevertheless, for the calculation of the bound-state
$\ell^-$--decay rate we will analyse the contribution of all excited
$nLM_L$--states of hydrogen ${\rm H}_{\ell}$, where $M_L = 0, \pm 1,
\ldots, \pm L$ is the magnetic quantum number.  We will show that the
main contribution comes from the $ns$--states only.

The wave function of  hydrogen  ${\rm H}_{\ell}$ in the $nLM_L$--state
we take in the form \cite{IV2}--\cite{IV6}
\begin{eqnarray}\label{label18}
\hspace{-0.3in}&&|{\rm H}^{(nLM_L)}_{\ell}(\vec{q}\,)\rangle =
 \frac{1}{(2\pi)^3}\sqrt{2 E_{{\rm H}^{(n)}_{\ell}}(\vec{q}\,)}\nonumber\\
\hspace{-0.3in}&&\times\int
 \frac{d^3k_{\ell}}{\sqrt{2E_{\ell}(\vec{k}_{\ell})}}
 \frac{d^3k_p}{\sqrt{2E_p(\vec{k}_p)}} \delta^{(3)}(\vec{q} -
 \vec{k}_{\ell} - \vec{k}_p)\nonumber\\
\hspace{-0.3in}&&\times\, \phi_{nLM_L}\Big(\frac{m_p \vec{k}_{\ell} -
 m_{\ell} \vec{k}_p}{m_p +
 m_{\ell}}\Big)\nonumber\\
\hspace{-0.3in}&&\times\,a^{\dagger}_{nLM_L}(\vec{k}_{\ell},\sigma_{\ell})
\, a^{\dagger}_p(\vec{k}_p,\sigma_p)|0\rangle,
\end{eqnarray}
where $E_{{\rm H}^{(n)}_{\ell}}(\vec{q}\,) = \sqrt{{M^{\;2}_{{\rm
H}^{(n)}_{\ell}}} + \vec{q}^{\;2}}$ and $\vec{q}$ are the energy and
the momentum of hydrogen, $M_{{\rm H}^{(n)}_{\ell}} = m_p + m_{\ell} +
\epsilon_n$ is the mass of hydrogen and $\epsilon_n$ is the binding
energy of the $nL$--state, $\phi_{nL M_L}(\vec{k}\,)$ is the wave
function of the $nLM_L$--state in the momentum representation \cite{BS57}
(see also \cite{IV2}--\cite{IV6}).

For the amplitude of the bound-state $\ell^-$--decay we obtain the
following expression
\begin{eqnarray}\label{label19}
\hspace{-0.3in}&&M({\rm H}^{(1s)}_m \to {\rm H}^{(nLM_L)}_{\ell} +
\tilde{\nu}_{\ell}) =\nonumber\\
\hspace{-0.3in}&&= -\,\sqrt{2} G_F V_{uq}\sqrt{\frac{2 M_{{\rm
H}^{(1s)}_m} 2 E_{{\rm H}^{(n)}_{\ell}}(\vec{q}\,) E_{\nu}}{2
M_m}}\,\delta_{\sigma_p \sigma}\nonumber\\
\hspace{-0.3in}&&\times
\int\frac{d^3k}{(2\pi)^3}\,\phi^*_{nLM_L}(\vec{k} -
\vec{k}_{\nu})\,\Phi_{1s}(\vec{k}\,)\nonumber\\
\hspace{-0.3in}&&\times\,\langle 0|A^{a +
ib}_{\mu}(0)|m(\vec{k}\,)\rangle\,[\varphi^{\dagger}_{\sigma_{\ell}}
\sigma^{\mu}\chi_{\tilde{\nu}_{\ell}}],
\end{eqnarray}
where $\varphi^{\dagger}_{\sigma_{\ell}}$ and
$\chi_{\tilde{\nu}_{\ell}}$ are the spinorial wave functions of the
lepton $\ell^-$ and the anti--neutrino $\tilde{\nu}_{\ell}$,
respectively, and $\sigma^{\mu} = (1, -\,\vec{\sigma}\,)$. The
contribution of the matrix element of the hadronic current we define
only by the {\it disconnected} part (see Appendix A).

Using the standard expression for the matrix element of the
axial--vector current \cite{PDG06,CA68}
\begin{eqnarray}\label{label20}
\langle 0|A^{a +
ib}_{\mu}(0)|m(\vec{k}\,)\rangle = i\sqrt{2}\,F_m\,Q_{\mu}
\end{eqnarray}
with $Q_{\mu} = (E_m, -\,\vec{k}\,)$, we get
\begin{eqnarray}\label{label21}
\hspace{-0.3in}&&M({\rm H}^{(1s)}_m \to {\rm H}^{(nLM_L)}_{\ell} +
\tilde{\nu}_{\ell}) =\nonumber\\
\hspace{-0.3in}&&= -\,2 i\, G_F V_{uq}F_m\sqrt{\frac{2 M_{{\rm
H}^{(1s)}_m}\, 2 E_{{\rm H}^{(n)}_{\ell}}(\vec{q}\,)E_{\nu}}{2
M_m}}\,\delta_{\sigma\sigma_p }\nonumber\\
\hspace{-0.3in}&&\times\,\Big\{M_m
[\varphi^{\dagger}_{\sigma_{\ell}}\, \chi_{\tilde{\nu}_{\ell}}] \int
d^3x\,\psi^*_{nLM_L}(\vec{r}\,)\Psi_{1s}(\vec{r}\,)\,e^{\textstyle
\,-\,i \vec{k}_{\nu}\cdot \vec{r}}\nonumber\\
\hspace{-0.3in}&&- i[\varphi^{\dagger}_{\sigma_{\ell}}\,\vec{\sigma}\,
\chi_{\tilde{\nu}_{\ell}}]\cdot \int
d^3x\,\psi^*_{nLM_L}(\vec{r}\,)\vec{\bigtriangledown}\,
\Psi_{1s}(\vec{r}\,)\nonumber\\
\hspace{-0.3in}&&\times\,e^{\textstyle \,-\,i\vec{k}_{\nu}\cdot
\vec{r}}\,\Big\},
\end{eqnarray}
where we have proceeded to the coordinate representation
\cite{Ivanov1}(see also \cite{Faber1}) and used the obvious
approximation $E_m \simeq M_m = m_m + \epsilon_{1s}$, as the main
contribution to the integral over $\vec{k}$ is defined by the region
$|\vec{k}\,|\sim 1/a_B = \alpha \mu_m$. Since de Broglie wave length
of the anti--neutrino is smaller compared to the Bohr radius of mesic
hydrogen, the main contribution to the spatial integral comes from the
origin. This gives
\begin{eqnarray}\label{label22}
\hspace{-0.3in}&&M({\rm H}^{(1s)}_m \to {\rm H}^{(nLM_L)}_{\ell} +
\tilde{\nu}_{\ell}) = -\,i\, G_F V_{uq}F_m\,\delta_{
\sigma\sigma_p}\nonumber\\
\hspace{-0.3in}&&\times\,\delta_{\sigma_{\ell},+\frac{1}{2}}\,\delta_{L0}\,
\delta_{M_L0}\,\sqrt{2 M_{{\rm H}^{(1s)}_m}\, 2 E_{{\rm
H}^{(n)}_{\ell}}(\vec{q}\,)\, 2 M_m E_{\nu}}\nonumber\\
\hspace{-0.3in}&&\times\,\psi_{n00}(0)\, \Big(1 -
\frac{E_{\nu}}{M_m}\Big)\,\Phi_{1s}(E_{\nu}),\nonumber\\
\hspace{-0.3in}&&
\end{eqnarray}
where $\psi_{n00}(0)$, the wave function of hydrogen ${\rm H}_{\ell}$
in the $ns$--state at the origin, is equal to $\psi_{n00}(0) =
\sqrt{\alpha^3\mu^3_{\ell}/\pi n^3}$ with the reduced mass of the
$\ell^-p$ pair $\mu_{\ell} = m_{\ell}m_p/(m_p + m_{\ell})$. This
testifies that the main contributions to the bound-state
$\ell^-$--decay rate of mesic hydrogen come from the $(ns)$--states,
which are splitted into the hyperfine $(ns)_F$--states with $F = 0$
and $F = 1$, respectively.

The bound-state $\ell^-$--decay rate of mesic hydrogen is defined by
\cite{Faber1}
\begin{eqnarray}\label{label23}
\hspace{-0.3in}&&\lambda_{\ell^-_b} = \sum^{\infty}_{n = 1}\sum_{F =
0,1}\lambda_{\ell^-_b}((ns)_F) = \frac{1}{2 M_{{\rm
H}^{(1s)}_m}}\sum^{\infty}_{n = 1}\nonumber\\
\hspace{-0.3in}&&\times \int \frac{d^3q}{(2\pi)^3 2 E_{{\rm
H}^{(n)}_{\ell}}}\frac{d^3k_{\nu}}{(2\pi)^3 2 E_{\nu}}\,(2\pi)^4
\delta^{(4)}(q + k_{\nu} - p)\nonumber\\
\hspace{-0.3in}&&\times
\,\frac{1}{2}\sum_{\sigma,\sigma_p,\sigma_{\ell}}|M({\rm H}^{(1s)}_m
\to {\rm H}^{(ns)}_{\ell} + \tilde{\nu}_{\ell})|^2.
\end{eqnarray}
The calculation of the r.h.s. of Eq.(\ref{label23}) is rather
straightforward and the result is
\begin{eqnarray}\label{label24}
\hspace{-0.3in}&&\lambda_{\ell^-_b} = \zeta(3)\,G^2_F|V_{uq}|^2 F^2_m
m_m \sqrt{(m_p + m_{\ell})^2 + E^2_{\nu}}\nonumber\\
\hspace{-0.3in}&&\times\,\frac{\alpha^3}{\pi^2}\,\Big(\frac{m_p
m_{\ell}}{m_p + m_{\ell}}\Big)^3\,\Big(1 -
\frac{E_{\nu}}{m_m}\Big)^2|\Phi_{1s}(E_{\nu})|^2\nonumber\\
\hspace{-0.3in}&&\times\,\frac{E^2_{\nu}}{(m_p + m_m)},
\end{eqnarray}
where $\zeta(3) = 1.202$ is the Riemann function, appearing as a
result of the summation over the {\it principal} quantum number $n$,
and $E_{\nu}$ is equal to
\begin{eqnarray}\label{label25}
E_{\nu} = (m_m - m_{\ell})\Big(1 - \frac{1}{2}\,
\frac{m_m - m_{\ell}}{m_p + m_m}\Big).
\end{eqnarray}
For the calculation of $\lambda_{\ell^-_b}$ we have neglected the
contribution of the binding energies. The numerical values of the
lifetimes of mesic hydrogen, caused by the bound-state
$\ell^-$--decays, are adduced in Table\,1.
{\renewcommand{\arraystretch}{1.5}
\renewcommand{\tabcolsep}{0.2cm}
\begin{table}[h]
\begin{tabular}{|l|c|c|}
\hline $\tau_{\ell^-_{b_{\phantom{x}}}}$ & ${\rm H}^{(1s)}_{\pi^-}$&
${\rm H}^{(1s)}_{K^-}$ \\ \hline $\tau_{\mu^-_{b_{\phantom{x}}}}$ &
15.83\,{\rm min} & 4.18\,{\rm yr}\\ \hline
$\tau_{e^-_{b_{\phantom{x}}}}$ & $2.29\times 10^8\,{\rm yr}$ &
$1.10\times 10^9\,{\rm yr}$\\ \hline
\end{tabular}
\caption{The lifetimes of pionic and kaonic hydrogen, caused by the
bound-state $\ell^-$--decays. The lifetimes are related to the decay
rates as $\tau_{\ell^-_b} = 1/\lambda_{\ell^-_b}$.}
\end{table}

\section{Concluding discussion}

We have calculated the continuum- and bound-state $\ell^-$--decay
rates of pionic and kaonic hydrogen in the $1s$ ground state, where
$\ell^-$ is either the electron $e^-$ or the muon $\mu^-$. According
to classification of $\beta$--decays \cite{EK66,HS66}, these are
allowed decays obeying the selection rule $\Delta J^P = 0^+$. To the
calculation of these decays we have applied the technique, which we
have used for the analysis of the H--like and He--like heavy
${^{140}}{\rm Pr}^{58+}$ and ${^{140}}{\rm Pr}^{57+}$ ions
\cite{Ivanov1} and the continuum- and bound-state $\beta^-$--decays of
bare ${^{207}}{\rm Tl}^{81+}$ and ${^{205}}{\rm Hg}^{80+}$ ions
\cite{Faber1}. The advantage of the $\ell^-$--decays of pionic and
kaonic hydrogen is that the hadronic matrix elements, equivalent to
the nuclear matrix elements for the weak decays of heavy ions, can be
calculated explicitly within current algebra and the PCAC hypothesis
for the $\pi^-$ and $K^-$ mesons \cite{CA68}.

We have shown that the decay rates of the continuum-state
$\ell^-$--decays of pionic and kaonic hydrogen practically coincide
with the decay rates of the free $\pi^-$ and $K^-$ meson. This means
that the Coulomb interaction, responsible for the existence of pionic
and kaonic hydrogen, is not enough to influence considerable on the
lifetime time of bound mesons. As a result the lifetimes of pionic and
kaonic hydrogen should be restricted by the lifetimes of the free
$\pi^-$ and $K^-$ mesons. The Coulomb corrections to the
continuum--state $\ell^-$--decay rate, caused by the Coulomb
interaction between the proton and lepton $\ell^-$ in the final
$p\ell^-\tilde{\nu}_{\ell}$ state, can be described by the Fermi
function \cite{EK66,HS66}.  However, for the continuum-state
$\ell^-$--decay of mesic hydrogen the final--state Coulomb interaction
between the proton and the lepton $\ell^-$ can be neglected, since the
main contribution to the integrals over the phase volume of the final
$p\ell^-\tilde{\nu}_{\ell}$ state comes from the region $E_{\ell} \sim
M_m/2$, where leptons are relativistic.

The lifetimes of pionic and kaonic hydrogen, caused by the bound-state
$\mu^-$--decays, are $\tau^{\pi^-}_{\mu^-_b} = 15.83\,{\rm min}$ and
$\tau^{K^-}_{\mu^-_b} = 4.18\,{\rm yr}$, respectively. In turn, the
lifetimes of pionic and kaonic hydrogen, related to the bound-state
$\beta^-$--decays, are of order of $\tau^{\pi^-}_{e^-_b} \sim
10^8\,{\rm yr}$ and $\tau^{K^-}_{e^-_b} \sim 10^9\,{\rm yr}$,
respectively.

\section*{Appendix A: Connected part of the  matrix element of the hadronic
 current} \renewcommand{\theequation}{A-\arabic{equation}}
 \setcounter{equation}{0}

The calculation of the {\it connected} part of the matrix element of
hadronic current we perform by using the reduction technique
\cite{BD65} and the PCAC hypothesis \cite{CA68,IV5}. This gives
\begin{eqnarray}\label{labelA.1}
\hspace{-0.3in}&&\langle
p(\vec{k}_p,\sigma_p)|J^{a+ib}_{\mu}(0)|m(\vec{k}\,)p(-\vec{k},\sigma)
\rangle_{\rm conn.} = \nonumber\\
\hspace{-0.3in}&&= \lim_{k^2 \to m^2_m}\frac{m^2_m -
k^2}{\sqrt{2}\,F_m m^2_m}(- i)\int d^4x\,e^{\textstyle\,-ik\cdot
x}\nonumber\\
\hspace{-0.3in}&&\times\,\langle
p(\vec{k}_p,\sigma_p)|T(A^{a+ib}_{\mu}(0)\partial^{\alpha}A^{a -
ib}_{\alpha}(x))|p(-\vec{k},\sigma)\rangle,\nonumber\\
\hspace{-0.3in}&&
\end{eqnarray}
where $T$ is a time--ordering operator \cite{BD65}. The calculation of
the r.h.s of Eq.(\ref{labelA.1}) we carry out in the chiral limit $k
\to 0$ ( soft--meson limit) and current algebra technique
\cite{CA68,IV5}. Using the relation
\begin{eqnarray}\label{labelA.2}
\hspace{-0.3in}&&T(A^{a+ib}_{\mu}(0)\partial^{\alpha}A^{a - ib}_{\alpha}(x)) =
\partial^{\alpha}T(A^{a+ib}_{\mu}(0)A^{a - ib}_{\alpha}(x))\nonumber\\
\hspace{-0.3in}&& - \delta(x^0)[A^{a+ib}_{\mu}(0),A^{a - ib}_0(x)],
\end{eqnarray}
taking the chiral limit and keeping only the leading contributions we
arrive at the following expression \cite{CA68,IV5}
\begin{eqnarray}\label{labelA.3}
\hspace{-0.3in}&&\langle
p(\vec{k}_p,\sigma_p)|J^{a+ib}_{\mu}(0)|m(\vec{k}\,)p(- \vec{k},\sigma)
\rangle_{\rm conn.} = \frac{i}{\sqrt{2}\,F_m }\nonumber\\
\hspace{-0.3in}&&\times\langle
p(\vec{k}_p,\sigma_p)|[A^{a+ib}_{\mu}(0),Q^{a -
ib}_5(0)]|p(\vec{0},\sigma)\rangle,
\end{eqnarray}
where $Q^{a - ib}_5(0)$ is axial charge operator \cite{CA68,IV5}. The
result of the calculation of the matrix element is
\cite{CA68,IV5,IV7}
\begin{eqnarray}\label{labelA.4}
\hspace{-0.3in}&&\langle
p(\vec{k}_p,\sigma_p)|J^{a+ib}_{\mu}(0)|m(\vec{k}\,)p(-
\vec{k},\sigma) \rangle_{\rm conn.} = \nonumber\\
\hspace{-0.3in}&& = \frac{i(k_p + p)_{\mu}}{\sqrt{2}\,F_m }\,{\cal
F}_m(\vec{k}^{\,2}_p),
\end{eqnarray}
where $k_p + p = (E_p(\vec{k}_p) + m_p, \vec{k}_p)$ and ${\cal
F}_m(\vec{k}^{\,2}_p)$ is the form factor defined by ${\cal
F}_{K^-}(\vec{k}^{\,2}_p) = 2{\cal F}_{\pi^-}(\vec{k}^{\,2}_p) = 2
{\cal F}_V(\vec{k}^{\,2}_p)$ \cite{IV5,IV7}.

The {\it connected} part of the matrix element of the hadronic
current is proportional to the amplitude of low--energy $mp$
scattering. Indeed, taking the limit $\vec{k}_p \to 0$ and multiplying
both sides of Eq.(\ref{labelA.4}) by the factor $(-i)
m_m/\sqrt{2}F_m$, related to the $m$--field in the final state
\cite{CA68}, we arrive at the Weinberg--Tomozawa term of $m$
scattering at threshold \cite{IV5,HP73}
\begin{eqnarray}\label{labelA.5}
\hspace{-0.3in}&&M(m p \to m p) = -
i\frac{m_m}{\sqrt{2}F_m}\,\langle p|J^{a+ib}_{\mu}(0)|m
p\rangle_{\rm conn.} = \nonumber\\ \hspace{-0.3in}&&= \frac{m_p
m_m}{F^2_m}\,{\cal F}_m(0),
\end{eqnarray}
where ${\cal F}_{\pi^-}(0) = 1$ and ${\cal F}_{K^-}(0) = 2$ for
$\pi^-p$ and $K^-p$ scattering (see \cite{HP73} and \cite{IV7}),
respectively.

For the continuum-state $\ell^-$--decay rate, caused by the {\it
connected} part of the matrix element of hadronic current, we get the
following expression
\begin{eqnarray}\label{labelA.6}
\lambda^{\rm (\it conn.)}_{\ell^-_c} &=& |\Psi_{1s}(0)|^2
\frac{G^2_F|V_{uq}|^2}{32\pi^3} \frac{m^2_{\ell}}{F^2_m} \frac{(M^2_m
  - m^2_{\ell})^2}{M^5_m}\nonumber\\ &&\times\int^{\infty}_0
dk_pk^2_p{\cal F}^2_m(k^2_p),
\end{eqnarray}
where $k_p = |\vec{k}_p|$ and $\Psi_{1s}(0) =
\sqrt{\alpha^3\mu^3_m/\pi}$ is the coordinate wave function of mesic
hydrogen in the ground state at the origin.

The ratio $R_{\ell^-_c} = \lambda^{\rm (\it
conn.)}_{\ell^-_c}/\lambda^{(m)}_{\ell^-_c}$ is equal
to
\begin{eqnarray}\label{labelA.7}
&&R^{(m)}_{\ell^-_c} = \frac{1}{8\pi^3}\frac{\alpha^3 \mu^3_m}{F^4_m
M^2_m}\int^{\infty}_0 dk_p k^2_p{\cal F}^2(k^2_p) =\nonumber\\ &&=
\frac{\alpha^3\mu^3_m M^3_V}{256\pi^2 M^2_m F^4_m}\,{\cal F}^2_m(0),
\end{eqnarray}
where $M_V = 843\,{\rm MeV}$ is a slop parameter \cite{IV7}. The
numerical values of the ratio for the $\pi^-$ and $K^-$ mesons is to
\begin{eqnarray}\label{labelA.8}
R^{(m)}_{\ell^-_c} =
\left\{\begin{array}{r@{\quad,\quad}l}
1.2\times 10^{-7} & m = \pi^-\\
3.2\times 10^{-7} & m = K^-.
\end{array}\right.
\end{eqnarray}
This value gives a hint that the contribution of the {\it connected}
part of the matrix element of the hadronic current to the
continuum-state $\ell^-$--decay rate should be small. The correct
contribution of the {\it connected} part to the continuum-state
$\ell^-$--decay rate one can estimate from the interference of the
{\it connected} and {\it disconnected} part of the matrix elements of
the hadronic current.  Using the result Eq.(\ref{labelA.7}) one finds
that the ratio of the continuum-state $\ell^-$--decay rate to the
interference term should be of order $0.1\,\%$. This shows that the
main contribution to the continuum-state $\ell^-$--decay rate of mesic
hydrogen is defined by the {\it disconnected} part of the matrix
element of hadronic current. Of course, the same conclusion is valid
for the bound-state $\ell^-$--decay rate.

\end{document}